\documentclass[sigconf]{acmart}
\AtBeginDocument{%
  }

\acmConference[EASE 2026]{The 30th International Conference on Evaluation and Assessment in Software Engineering}{9–12 June, 2026}{Glasgow, Scotland, United Kingdom}
\usepackage{xcolor}
\usepackage{listings}
\usepackage{multirow}

\definecolor{green}{RGB}{64, 160 ,43}
\definecolor{ctpKeyword}{rgb}{0.92, 0.46, 0.80}
\definecolor{ctpThis}{rgb}{0.82, 0.06, 0.22}
\definecolor{ctpMethod}{rgb}{0.12, 0.40, 0.96}
\definecolor{ctpComment}{rgb}{0.36, 0.37, 0.47}
\definecolor{ctpText}{rgb}{0.18, 0.20, 0.27}

\lstdefinelanguage{TypeScript}{
  keywords={typeof, new, true, false, catch, function, return, null, catch, switch, var, if, in, while, do, else, case, break, interface, const, as, any, let, string, number, boolean, enum, undefined, type, never},
  keywordstyle=\color{ctpKeyword}\bfseries,
  ndkeywords={string, number, boolean, undefined, any},
  ndkeywordstyle=\color{ctpMethod},
  identifierstyle=\color{ctpText}\ttfamily\bfseries,
  sensitive=true,
  comment=[l]{//},
  morecomment=[s]{/*}{*/},
  commentstyle=\color{ctpComment}\ttfamily\itshape,
  stringstyle=\color{green}\ttfamily\bfseries,
  morestring=[b]',
  morestring=[b]"
}

\begin{document}

\title{An Empirical Evaluation of Code Smell Detection in Angular Applications}


\author{Maykon Nunes}
 \normalsize \email{maykon.nunes@alu.ufc.br}
\orcid{0009-0003-5442-7413}
\affiliation{%
  \normalsize  \institution{Federal University of Ceará}
  \city{Quixadá}
  \country{Brazil}
}

\author{Emanuel Coutinho}
 \normalsize \email{emanuel.coutinho@ufc.br}
\orcid{0000-0003-2233-7109}
\affiliation{%
  \normalsize  \institution{Federal University of Ceará}
  \city{Quixadá}
  \country{Brazil}
}

\author{Carla Bezerra}
 \normalsize \email{carlailane@ufc.br}
\orcid{0000-0002-5879-5067}
\affiliation{%
  \normalsize  \institution{Federal University of Ceará}
  \city{Quixadá}
  \country{Brazil}
}

\author{Ivan Machado}
\normalsize \email{ivan.machado@ufba.br}
\orcid{0000-0001-9027-2293}
\affiliation{%
  \institution{Federal University of Bahia}
  \city{Salvador}
  \country{Brazil}
}

\renewcommand{\shortauthors}{Trovato et al.}

\begin{abstract}
  Angular is one of the most widely adopted frameworks for developing large-scale, dynamic web applications. As projects increase in scope and complexity, developers face growing challenges in managing architecture and maintaining clean, modular code. These challenges often lead to design flaws, commonly referred to as code smells. While React-specific smells have been cataloged in prior studies, limited knowledge exists regarding Angular-specific smells and how they manifest. This study investigates Angular code smells through a grey literature review, consolidating community knowledge and technical discussions. From the collected sources, 11 distinct Angular code smells were identified, 6 of which also occur in React-based systems, suggesting that some issues are cross-framework. Each smell was analyzed, exemplified, and grouped according to its technical characteristics. Based on the resulting catalog, we implemented an automated static analysis tool to detect Angular code smells. The tool was empirically evaluated using a manually validated dataset, and its effectiveness was assessed through standard information retrieval metrics. The evaluation results indicate high detection performance across all smells, achieving accuracy values above 0.88 and F1-scores ranging from 0.89 to 1.00. The findings reveal recurring issues such as component overloading, duplicated logic, and inefficient template bindings, reinforcing the relevance of systematic detection support. This study presents the first catalog of Angular-specific code smells derived from grey literature and demonstrates the feasibility and effectiveness of automated detection, providing a solid foundation for future empirical studies and tool development aimed at improving front-end code quality.
\end{abstract}

\begin{CCSXML}
<ccs2012>
<concept>
<concept_id>10011007.10011006.10011072</concept_id>
<concept_desc>Software and its engineering~Software libraries and repositories</concept_desc>
<concept_significance>500</concept_significance>
</concept>
<concept>
<concept_id>10011007.10011006.10011050</concept_id>
<concept_desc>Software and its engineering~Context specific languages</concept_desc>
<concept_significance>500</concept_significance>
</concept>
</ccs2012>
\end{CCSXML}

\ccsdesc[500]{Software and its engineering~Software libraries and repositories}
\ccsdesc[500]{Software and its engineering~Context specific languages}

\keywords{Angular, Code Smells, Software Design, Web Development}


\maketitle

\section{Introduction}

JavaScript’s front-end ecosystem has grown extensively, with frameworks like \textsc{React}, \textsc{Angular}, and \textsc{Vue} playing a central role in building modern web applications \cite{fabio2021Adoption, Novac2021ReactComparative, rajput2023comparing}. Libraries such as \textsc{React} become popular due to their performance, reusability, and component-based architecture~\cite{LorenzReactCOP}. However, these frameworks also introduced design issues that can compromise code quality \cite{Ferreira2023ReactSmells, maykonReactTypeScript}. 

Such design issues are manifested through code smells, a term introduced by \cite{fowler2018refactoring} to describe symptoms of bad quality code that may indicate deeper design problems. They have been extensively studied in object-oriented systems \cite{lacerda2020codesmells, zakeri2023datasets, Sharma2017SmellsCSharp}. Nevertheless, front-end frameworks such as \textsc{Angular} remain underexplored, both in terms of empirical evidence and tools.

Despite its widespread adoption, the framework provides a robust infrastructure for developing scalable applications through modules, components, and dependency injection mechanisms~\cite{Enes2019AngularComponentBased}. Moreover, \textsc{Angular} integrates several features such as two-way data binding, template-based syntax, and RESTful API handling~\cite{Slavina2019AngularEducation, Novac2025AngularBlazor, Geetha2022AngularInterpretation}. While these characteristics enable the construction of Single-Page Applications~\cite{Shahoor2025SPAsMemoryLeakage}, they also increase the architectural and design complexity of Angular projects, potentially leading to code smells.

Previous studies have investigated the presence of code smells in front-end technologies such as plain JavaScript~\cite{lacerda2020codesmells, Fard2013JsNose}, as well as in frameworks and libraries like \textsc{React}~\cite{Ferreira2023ReactSmells} and combining \textsc{React} with \textsc{TypeScript}~\cite{maykonReactTypeScript}. However, research addressing code smells specific to this framework remains scarce. To the best of our knowledge, this is the first work that proposes a catalog of \textsc{Angular}-specific code smells, grounded on evidence from grey literature and implementation in a detection tool.

In this study, we conducted a grey literature review to identify and classify code smells relevant to \textsc{Angular}, following a systematic process. We collected and analyzed content published by practitioners with practical experience in \textsc{Angular} development. This investigation resulted in a set of recurring design issues that reflect common quality concerns in the framework. Each smell was characterized according to its manifestation and underlying design problem. In addition, we developed a static analysis tool that extends the detection capabilities of existing smell detectors (e.g., \textsc{ReactSniffer} and \textsc{SniffTSX}), offering unified and extensible support for multiple front-end frameworks.

These results motivate the main contributions of this study: (i) a catalogue of \textsc{Angular}-specific code smells derived from a systematic analysis of practitioner discussions in grey literature; (ii) a comparative investigation highlighting overlaps between \textsc{Angular} and \textsc{React} smells, suggesting opportunities for cross-framework generalization; and (iii) the design and implementation of an extensible static analysis tool that integrates existing detectors to automatically identify framework-specific code smells across technologies.

\section{Background}

\textsc{Angular} is an open-source web application framework maintained by Google and the developer community. It was designed to facilitate the development of dynamic and scalable single-page applications (SPAs).  Unlike libraries such as \textsc{React}, Angular provides a full-featured framework that includes routing, form management, HTTP communication, and dependency injection.

\textsc{Angular} applications are typically written in TypeScript, a statically typed superset of JavaScript that enhances reliability and tooling support. The framework uses decorators to define metadata for its main building blocks. For instance, the \texttt{@Component} decorator marks a class as a UI component, associating it with a specific HTML template and styles (see Listing~\ref{lst:example-angular-component}). The template defines the structure of the user interface using HTML combined with \textsc{Angular’s} declarative syntax.

\begin{lstlisting}[caption=Example of Angular component, label={lst:example-angular-component}]
@Component({
    selector: 'app-user',
    template: '<p>Name: {{ name }}</p>'
})
export class UserComponent {
    name: string = 'Bob';
}
\end{lstlisting}

\textsc{Angular} adopts a dependency injection mechanism to support modular and maintainable application design. Shared functionalities are organized into services, which are classes annotated with the \texttt{@Injectable} decorator. When injected into a component or another service, they provide access to the encapsulated business logic and utility methods defined within the service.

\begin{lstlisting}[caption=Example of service injection, label={lst:angular-service}]
@Injectable({ providedIn: 'root' })
export class UserService {
    getName(): string { 
        return 'Bob'; 
    }
}

@Component({ 
    selector: 'app-user', 
    template: '{{ user }}' 
})
export class UserComponent {
    name: string;
    
    constructor(private userService: UserService) {
        this.name = this.userService.getName();
    }
}
\end{lstlisting}

Another important feature of \textsc{Angular} is the two-way data binding mechanism, which enables automatic synchronization between the model and the view layers. Therefore, changes in component properties are immediately reflected in the corresponding user interface elements, while user interactions in the interface automatically update the component state. For example, a text input can be bound to a component property using the \texttt{[(ngModel)]} directive, ensuring that updates in the input field are reflected in the property (see Listing~\ref{lst:two-way-binding}.

\begin{lstlisting}[caption=Example of two-way data binding, label={lst:two-way-binding}]
<input [(ngModel)]="name" placeholder="Enter name">
<p>Hello, {{ name }}!</p>
\end{lstlisting}

This interaction between the component and the view is defined through templates, which provide mechanisms to dynamically display and manipulate data. Interpolation, denoted by double curly braces \texttt{\{\{ ... \}\}}, allows component properties to be rendered directly within HTML elements. Additionally, directives extend standard HTML elements with additional functionality: structural directives, such as \texttt{*ngIf} and \texttt{*ngFor}, modify the DOM by conditionally including or repeating elements, while attribute directives adjust their presentation or behavior. An example combining these features is shown in Listing~\ref{lst:angular-directives}, where a list of users is rendered conditionally based on a component property.

\begin{lstlisting}[caption=Example of directives and interpolation, label={lst:angular-directives}]
<ul *ngIf="users.length > 0; else emptyList"> 
    <li *ngFor="let user of users">{{ user.name }}</li> 
</ul>
<ng-template #emptyList>
    <p>No users available.</p> 
</ng-template>
\end{lstlisting}

In addition to the interaction between components and templates, \textsc{Angular} applications are developed using \textsc{TypeScript} as the primary programming language. \textsc{TypeScript}, a statically typed superset of JavaScript, introduces compile-time type checking and language constructs such as interfaces, generics, and decorators. These features contribute to the definition of contracts between components and their data models. Listing~\ref{lst:typescript-interface} illustrates the use of an interface to define the structure of a data object within a component.

\begin{lstlisting}[caption=Example of TypeScript interface usage, label={lst:typescript-interface}]
interface User {
    id: number;
    name: string;
}

@Component({
    selector: 'app-user',
    template: '<p>{{ user.name }}</p>'
})
export class UserComponent {
    user: User = { id: 1, name: 'Bob' };
}
\end{lstlisting}

\section{Study Design}\label{sec:study-design}

The objective of this study is to identify code smells for the Angular framework, assisting front-end developers in this framework to detect and refactor these anomalies. In this section, we describe the procedures used to prospect a catalog of \textsc{Angular} code smells.

\subsection{Research Questions}

Four research questions guide the present study:

\begin{itemize}
    \item[\textbf{RQ$_1$}] \textbf{What are the most common code smells when developing with \textsc{Angular}?} This RQ aims to identify and categorize the code smells mentioned in grey literature sources according to their technical nature and design characteristics.
    \item[\textbf{RQ$_2$}] \textbf{Where are \textsc{Angular} code smells discussed by the developer community?} The goal of this question is to investigate the types of grey literature sources in which \textsc{Angular} code smells are discussed.
    \item[\textbf{RQ$_3$}] \textbf{Which \textsc{React}-specific code smells are also discussed by \textsc{Angular} practitioners?} In this RQ, we examine the overlap between \textsc{Angular} and \textsc{React} code smells, comparing the identified \textsc{Angular} smells with existing React smell catalogs to determine which design issues are cross-framework.
    \item[\textbf{RQ$_4$}] \textbf{How can existing code smells tools for \textsc{React} be extended to support the detection of \textsc{Angular} code smells?} This RQ aims to investigate how \textsc{ReactSniffer} can be integrated with a newly developed \textsc{Angular} smell detector.
    \item[\textbf{RQ$_5$}] \textbf{How accurate and effective is the code smell detection tool in identifying code smells in open-source \textsc{Angular} projects?} This question aims to measure the accuracy of code smell detection in Angular using the developed tool.
\end{itemize}

\subsection{Grey Literature Review}\label{subsec:grey-literature-review}

\begin{figure}[ht!] 
    \centering
\includegraphics[width=\columnwidth]{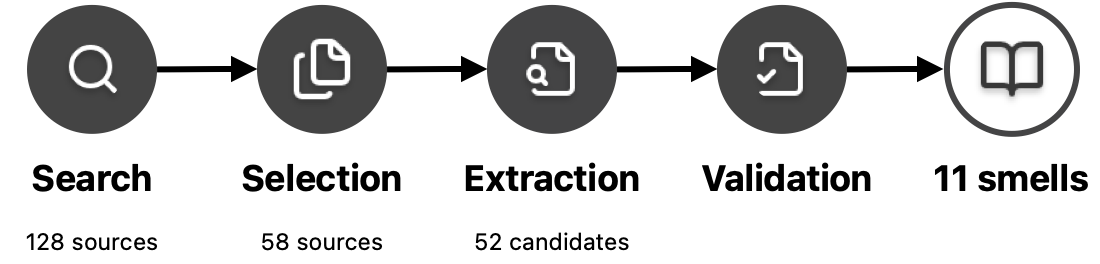}
    \caption{Methodology for cataloging code smells in Angular}
    \label{fig:grey-literature-review}
\end{figure}

Since few studies investigate quality issues on the framework, the grey literature, composed of forums, blogs, videos, etc., serves as a valuable source of practical knowledge \cite{Zhang2020GreyLiterature}. To achieve this, we follow the guidelines proposed by Garousi et al. \cite{Garousi2019GuidelinesGreyLiterature}. Figure \ref{fig:grey-literature-review} summarizes the steps we followed.

\subsubsection{Search}\label{subsubsec:search}

To support our investigation, we constructed a search query that included the ``Angular'' keyword followed by the synonyms of ``code smell'' in both singular and plural forms, combined using OR clauses, as in Listing~\ref{lst:search-query}. Then, we executed this query on Google in June 2025 and only considered sources written in English.

\begin{lstlisting}[caption=Search query, label={lst:search-query}]
("Angular") AND
("code smell" OR "code smells" OR "anti-pattern" OR "anti-patterns" OR "anti pattern" OR "anti patterns" OR "antipattern" OR "antipatterns" OR "bad smell" OR "bad smells" OR "bad-practice" OR "bad-practices" OR "bad practice" OR "bad practices")
\end{lstlisting}

According to Garousi et al. \cite{Garousi2019GuidelinesGreyLiterature}, GLRs require a stopping condition due to evidence exhaustion. In this work, the search was limited to the first 100 results, but could be extended if subsequent results continued to reveal relevant sources.

\subsubsection{Selection}\label{subsubsec:selection}

Since the GLR sources were not peer-reviewed, we derived a checklist based on the Quality Assessment Checklist to ensure that only sources meeting essential quality criteria were included. Furthermore, our questions were based on two aspects: (1) source credibility, considering affiliation with companies working with Angular; and (2) author expertise, including prior publications and recognized experience in Angular development. A source was considered acceptable if it met at least one of the listed criteria in Table \ref{tab:verification-list}.

\begin{table} \small
    \caption{Quality assessment checklist}
    \label{tab:verification-list}
    \begin{tabular}{p{2cm} p{5cm}}
         \toprule
            \textbf{Criteria} & \textbf{Questions} \\
         \midrule
            \multirow{2}{*}{Source credibility} & Is the publication linked to an organization working with Angular? \\
            ~ & Is the author associated with an organization that uses Angular? \\
         \midrule
            \multirow{2}{*}{Author expertise} & Has the author published other content about Angular? \\
            ~ & Does the author demonstrate expertise in Angular? \\
         \bottomrule
    \end{tabular}
\end{table}

We initially selected 128 sources. Then, these documents were read by two authors in order to exclude those that did not meet our quality criteria (26 sources), were not related to code smells or Angular (23 sources), were duplicates (10 sources), referred to the deprecated \textsc{AngularJS} framework (5 sources), or discussed only traditional code smells unrelated to Angular (5 sources). After this step, 58 sources were selected for analysis, which we refer to as S1 to S58.

\subsubsection{Data extraction and validation}\label{subsubsec:data-extraction-and-validation}

The sources identified in the previous step were carefully analyzed by the first author using thematic analysis to identify recurring patterns related to code smells and their impact on software quality attributes. The extracted data were organized in a tabular, where each row corresponds to a specific code smell along with its associated sources. 

The preliminary set of candidate smells was reviewed with the co-authors to ensure agreement. This process, grounded in the analysis of 57 sources, resulted in the identification of 52 candidate code smells, some of which were reported across multiple documents.

The validation process began with a detailed analysis conducted by the first author, a developer with experience with front-end frameworks, relying solely on the content explicitly provided in the sources, and was subsequently reviewed by the co-authors. When clarification of framework aspects was required, external references were consulted to support the interpretation. To ensure the relevance of the resulting set, we retained only those candidate smells that (i) matched a smell previously cataloged to React proposed by Ferreira and Valente \cite{Ferreira2023ReactSmells}, and (ii) appeared in at least two distinct sources.

Candidate smells that did not meet our inclusion criteria were typically associated with low-level concerns, often tied to specific implementation practices rather than broader design issues. For instance, the candidate smell \textsc{Nested Subscriptions} was excluded, as it pertains specifically to the misuse of the \textsc{RxJS} library, rather than representing a design-related smell inherent to \textsc{Angular}.

The candidate smells \textsc{Default Change-Detection-Strategy instead of On-Push}, \textsc{*No trackBy in ngFor}, and \textsc{Unoptimized Module Loading} were not included due to a lack of agreement among the authors. Although these practices may affect performance, their classification as code smells is highly dependent on project-specific factors, such as the need for frequent UI updates or architectural constraints in Micro Frontend scenarios, making them context-sensitive rather than universally indicative of poor design.

We conclude this step with 11 identified smells, summarized in Table~\ref{tab:identified-code-smells}.

\begin{table} \small
    \caption{Identified code smells}
    \label{tab:identified-code-smells}
    \begin{tabular}{p{4.5cm}ll}
         \toprule
            \textbf{Code smells} & \textbf{\textit{Sources}} & \textbf{\textit{\#}} \\
         \midrule
            Large Component & S2, S3, S17, S20, S24, S25 & 6 \\
            Inefficient method binding in templates & S1, S2, S4, S7, S23 & 5 \\    
            Overusing Any Type & S3, S24, S25, S32 & 4 \\
            Excessive Parent-to-Child Communication & S2, S7, S35 & 3 \\
            Direct DOM Manipulation & S3, S27, S35 & 3 \\
            Coupled Services & S3, S57 & 2 \\ 
            Inheritance Instead of Composition & S3, S12 & 2 \\
            Too Many Inputs & S3, S40 & 2 \\
            Prop Drilling & S9 & 1 \\
            Large File & S13 & 1 \\
            Duplicated Component & S20 & 1 \\
         \bottomrule
    \end{tabular}
\end{table}

\subsubsection{Code smells classification}\label{subsubsec:code-smells-classification}

We also classify these smells into two categories:

\begin{itemize}
    \item \textit{Angular-specific}: This category includes code smells that are inherent to the \textsc{Angular} ecosystem and its associated technologies, such as its use of dependency injection, change detection strategies, and template syntax.
    \item \textit{Cross-framework}: This category includes code smells originally identified in other front-end frameworks, particularly in \textsc{React}, as cataloged by \citet{Ferreira2023ReactSmells}.
\end{itemize}

\subsection{Tools Integration}\label{subsec:tools-integration}

The proposed tool constitutes an integration of existing static analysis frameworks, specifically \textsc{ReactSniffer} and \textsc{SniffTSX}. By integrating these analyzers, the tool extends its capabilities to identify code smells specific to \textsc{Angular}. 

\begin{figure}[ht!] 
    \centering
    \includegraphics[width=\columnwidth]{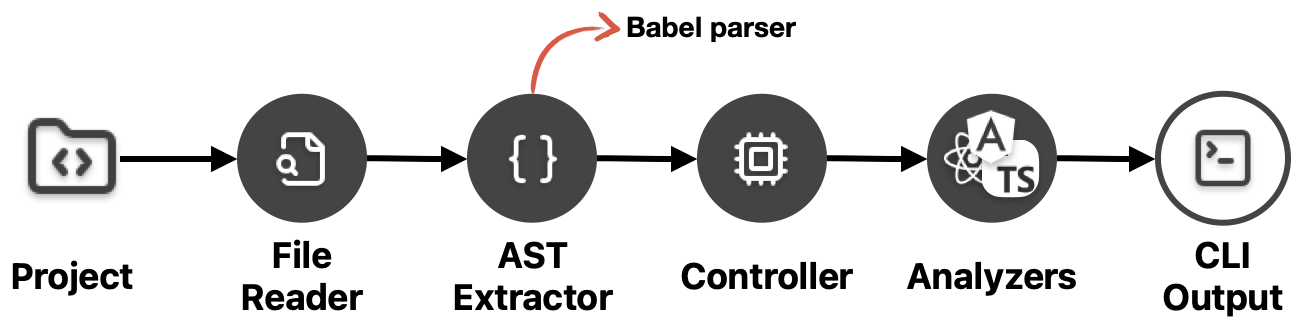}
    \Description[tool architecture]{A diagram representing the architecture of the tool. It includes modules responsible for code parsing, smell detection and reporting}
    \caption{Component diagram of the tool}
    \label{fig:tool-architecture}
\end{figure}

As Figure \ref{fig:tool-architecture} shows, the first step is performed by the \textit{File Reader}, which traverses the target project and retrieves all source files. This component filters files according to their extensions (e.g., \texttt{.js}, \texttt{.jsx}, \texttt{.ts}, \texttt{.tsx}) and project configuration files (e.g., \texttt{package.json}) required for determining the framework front-end employed in the project.

Once the framework has been identified, the retrieved files are passed to the \textit{AST Extractor}, implemented using \textsc{Babel} parser, which generates the abstract syntax tree (AST) for each file. Next, the \textit{Controller} coordinates the execution of the detection modules according to the identified framework. This module ensures that the \textsc{ReactSniffer} detector operates exclusively on \textsc{React} projects, while the \textsc{SniffTSX} addresses \textsc{TypeScript}-specific smells that can occur in both \textsc{React} and \textsc{Angular}.

After aggregating the results from the detectors, all identified code smells, along with their file locations and corresponding line numbers, are presented through the \textit{CLI Output}. The architecture remains unified and extensible, allowing the addition of new detectors and support for other front-end frameworks in future work.

\section{Results}\label{sec:results}

This section presents the results of the steps described in the previous section.

\subsection{(\textbf{RQ$_1$}) What are the most common code smells when developing with \textsc{Angular}?}\label{subsec:rq1}

The review revealed 11 code smells commonly discussed by Angular developers. For each, we present a description and a practical example.

\subsubsection{Inefficient method binding in templates}\label{subsubsec:inefficient-method-binding-in-templates}

\textsc{Angular} provides two main change detection strategies to manage how and when the view is updated in response to state changes: The \texttt{Default} strategy, which checks the entire component tree, and the \texttt{OnPush} strategy, which limits checks to components with changed inputs or triggered events.

\begin{lstlisting}[language=TypeScript]
<div>
    <p>Total: {{ calculateTotal() }}</p>
</div>
\end{lstlisting}

\begin{lstlisting}[language=TypeScript, caption=Angular component with Inefficient method binding in templates, label={lst:inefficient-method-binding-in-templates}]
@Component({
  selector: 'app-cart',
  templateUrl: './app-cart.component.html'
})
export class CartComponent {
  items = [{ price: 10 }, { price: 15 }, ...];

  calculateTotal(): number {
    return this.items.reduce((sum, item) => sum + item.price, 0);
  }
}
\end{lstlisting}

A commonly reported case by the practitioners involves binding methods directly in templates under the default strategy. In Listing~\ref{lst:inefficient-method-binding-in-templates}, \texttt{calculateTotal()} is re-evaluated every time \textsc{Angular} runs change detection, even if \texttt{item} is not changed. This happens because the framework does not cache the return value of the method, as it is not a pure function. Since Angular cannot determine whether the method is free of side effects or consistently returns the same output, it executes the method on every cycle to ensure correctness.

%From a design perspective, the presence of this smell reveals a partial misalignment with \textsc{Angular}’s declarative and reactive programming paradigms. Templates are intended to express data dependencies rather than perform imperative computations. Moreover, the use of the \texttt{OnPush} change detection strategy can alleviate the problem by restricting re-evaluation to components whose inputs have changed. However, it should be noted that this strategy alone does not eliminate the performance issue when non-pure methods remain bound to the template, as these methods continue to be executed upon each detected change.

\subsubsection{Overusing any type}\label{subsubsec:overusing-any-type}

In \textsc{Angular}, developers tend to use the type \texttt{any}, a type annotation that effectively disables type checking, allowing a variable to accept any value without restrictions. When a variable is declared with \texttt{any}, the \textsc{TypeScript} compiler bypasses type verification, providing maximum flexibility while eliminating the fundamental benefits of static typing.

Practitioners report that excessive use of the \texttt{any} type compromises type safety, one of \textsc{TypeScript}'s primary advantages, allowing type errors to pass undetected during compilation and manifest only at runtime. Additionally, extensive use of \texttt{any} can mask design and architectural issues, creating a false sense of simplicity that results in more fragile and error-prone code.

In Listing~\ref{lst:overusing-any-type}, the \texttt{UserProfileComponent} have multiple instances of \texttt{any}. The component declares the \texttt{user} property and \texttt{users} array with \texttt{any} types, eliminating compile-time type checking for these critical data structures. The \texttt{loadUser} method accepts an \texttt{any} parameter, which could lead to runtime errors if an invalid type is passed. Furthermore, the HTTP response from \texttt{userService.getUser} is typed as \texttt{any}, preventing the compiler from validating the structure of the received data. The \texttt{processUserData} method compounds the issue by accepting \texttt{any} input and returning \texttt{any} output, creating a cascade of type safety violations throughout the component's data flow.

\begin{lstlisting}[language=TypeScript, caption=Angular component with Overusing Any Type, label={lst:overusing-any-type}]
@Component({
  selector: 'app-user-profile',
  templateUrl: './app-user-profile.component.html'
})
export class UserProfileComponent {
  user: any;
  users: any[] = [];
  
  constructor(private userService: UserService) {}
  
  loadUser(id: any): void {
    this.userService.getUser(id).subscribe((data: any) => {
      this.user = data;
      this.processUserData(data);
    });
  }
}
\end{lstlisting}

\subsubsection{Excessive parent-to-child communication}\label{subsubsec:excessive-parent-to-child-communication}

Component-based architectures encourage a modular and decoupled structure where communication between components occurs through well-defined and constrained mechanisms. However, some developers report the excessive use of \texttt{@ViewChild}, a feature that allows a parent component to access a child component's public interface directly, as a source of architectural degradation.

This anti-pattern introduces a strong dependency between the parent and the internal implementation of the child. As a result, changes to the child component may propagate to the parent, as in Listing~\ref{lst:excessive-parent-to-child-communication}.

\begin{lstlisting}[language=TypeScript]
@Component({
  selector: 'app-child',
  template: `<p>{{ message }} - {{ count }}</p>`
})
export class ChildComponent {
  message = 'Init';
  count = 0;

  update(msg: string) {
    this.message = msg;
    this.count++;
  }

  reset() {
    this.message = 'Init';
    this.count = 0;
  }

  /* ... */
}
\end{lstlisting}

\begin{lstlisting}[language=TypeScript, caption=Angular component with Excessive Parent-to-Child Communication, label={lst:excessive-parent-to-child-communication}]
@Component({
  selector: 'app-parent',
  template: `
    <app-child></app-child>
    <button (click)="init()">Init</button>
    <button (click)="activate()">Activate</button>
    <button (click)="disable()">Disable</button>
    <button (click)="increment()">Increment</button>
    <button (click)="reset()">Reset</button>
  `
})
export class ParentComponent {
  @ViewChild(ChildComponent) child!: ChildComponent;

  updateChild() {
    this.child.update('Updated by parent');
  }

  resetChild() {
    this.child.reset();
  }

  /* ... */
}
\end{lstlisting}

The parent directly alters the child's state by invoking its method. This implementation compromises component independence and hinders unit testing, as child components can no longer operate in isolation. Moreover, this tight coupling may escalate over time, leading to rigid component hierarchies.

\subsubsection{Coupled Services}\label{subsubsec:coupled-services}

When a single service becomes responsible for handling multiple unrelated concerns, it introduces tight coupling. This violates the principle of Separation of Concerns and makes the service difficult to maintain, test, or reuse. In Listing~\ref{lst:coupled-service}, the \texttt{AppService} handles both user and item logic, and is injected into components with unrelated responsibilities.

\begin{lstlisting}[language=TypeScript, caption=Example of a tightly coupled service, label={lst:coupled-service}]
@Injectable({ providedIn: 'root' })
export class AppService {
  getUser() { /* ... */ }
  getItems() { /* ... */ }
}

@Component({ /* ... */ })
export class HeaderComponent {
  user = this.appService.getUser();
  constructor(private appService: AppService) {}
}

@Component({ /* ... */ })
export class ListComponent {
  items = this.appService.getItems();
  constructor(private appService: AppService) {}
}
\end{lstlisting}

From a design perspective, this smell represents a violation of the \textit{Single Responsibility Principle} and resembles the well-known \textit{God Class} anti-pattern at the service layer. While \textsc{Angular}'s dependency injection mechanism facilitates modularization, it can also mask this issue by allowing developers to inject a single service into multiple unrelated components. As a result, business logic that should reside in independent, cohesive services tends to accumulate in a centralized one, reducing separation of concerns.

\subsubsection{Large component}\label{subsubsec:large-component}

A large component refers to a component that contains excessive code, handles multiple responsibilities, or encompasses too many features, violating the Single Responsibility Principle. Practitioners advocate for component decomposition as a fundamental practice, in which complex or reusable user interface (UI) portions should be extracted into separate, focused components.

\begin{lstlisting}[language=TypeScript, caption=Large component handling multiple responsibilities, label={lst:large-component}]
@Component({
  selector: 'app-dashboard',
  templateUrl: './dashboard.component.html'
})
export class DashboardComponent {
  users: User[] = [];
  logs: string[] = [];
  chartData: number[] = [];

  constructor(private userService: UserService, private logService: LogService) {}

  ngOnInit() {
    this.loadUsers();
    this.loadLogs();
    this.generateChartData();
  }

  loadUsers() { /* ... */ }
  loadLogs() { /* ... */ }
  generateChartData() { /* ... */ }
}
\end{lstlisting}

In Listing~\ref{lst:large-component}, the \texttt{DashboardComponent} manages user data retrieval, system logging, and chart generation within a single class. This concentration of responsibilities increases cognitive complexity and reduces cohesion. Each functionality could instead be isolated into components.

\subsubsection{Direct DOM manipulation}\label{subsubsec:direct-dom-manipulation}

In web development, the Document Object Model (DOM) represents the structure of a web page as a tree of objects that can be programmatically accessed and modified. As usual in \textsc{Angular}, developers can access DOM elements through the \textsc{ViewChild} decorator and manipulate them using the \texttt{ElementRef} class. While this approach may seem convenient, it bypasses \textsc{Angular}'s declarative and reactive paradigms. In Listing~\ref{lst:direct-dom-manipulation}, the \texttt{ElementRef} is used to directly alter an element's style.

\begin{lstlisting}[language=TypeScript, caption=Component with Direct DOM Manipulation, label={lst:direct-dom-manipulation}]
@Component({
  selector: 'app-alert',
  template: `<div #alertBox>Alert message</div>`
})
export class AlertComponent {
  @ViewChild('alertBox') alertBox!: ElementRef;

  ngAfterViewInit() {
    this.alertBox.nativeElement.style.backgroundColor = 'red';
  }
}
\end{lstlisting}

\subsubsection{Inheritance instead of composition}\label{subsubsec:inheritance-instead-of-composition}

While Angular supports both inheritance and composition, some developers heavily rely on class inheritance to share logic across components, often by creating abstract base classes. However, this approach can introduce tight coupling between components and reduce maintainability. As illustrated in Listing~\ref{lst:inheritance-instead-of-composition}, a base class encapsulates shared functionality.

\begin{lstlisting}[language=TypeScript, caption=Component with Inheritance Instead of Composition, label={lst:inheritance-instead-of-composition}]
export abstract class BasePageComponent {
  abstract pageTitle: string;

  initPage() {
    console.log(`Initializing page: ${this.pageTitle}`);
  }
}

@Component({
  selector: 'app-home',
  templateUrl: './home.component.html'
})
export class HomeComponent extends BasePageComponent {
  pageTitle = 'Home Page';
}
\end{lstlisting}

Moreover, developers report that subclasses frequently require behavior not anticipated in the base class, leading to abstract methods tailored to specific cases. This results in logic being spread across the hierarchy, which hampers modularity, flexibility, and ease of testing.

\subsubsection{Too many inputs}\label{subsubsec:too-many-inputs}

The \texttt{@Input} decorator is used to define properties that allow a child component to receive data from its parent. However, relying on an excessive number of external inputs can be problematic. It often indicates that the component is overly complex or tightly coupled to its parent, which undermines modularity (see Listing~\ref{lst:too-many-inputs}).

\begin{lstlisting}[language=TypeScript, caption=Component with Too Many @Inputs, label={lst:too-many-inputs}]
@Component({
  selector: 'app-user-card',
  templateUrl: './user-card.component.html'
})
export class UserCardComponent {
  @Input() userName!: string;
  @Input() userAge!: number;
  @Input() userEmail!: string;
  @Input() userRole!: string;
  @Input() isActive!: boolean;
  @Input() showAvatar!: boolean;
  @Input() highlight!: boolean;
  /* ... */
}

\end{lstlisting}

\subsubsection{Prop drilling}\label{subsubsec:prop-drilling}

\textsc{Prop Drilling} arises when data must be passed from a parent component to a deeply nested child component through multiple intermediary components that do not use the data themselves. This smell typically occurs through successive declarations of \texttt{@Inputs} properties, resulting in tightly coupled component hierarchies and reduced maintainability.

\begin{lstlisting}[language=TypeScript, caption=Prop Drilling across multiple nested components, label={lst:prop-drilling}]
@Component({
  selector: 'app-parent',
  template: `<app-child-a [data]="info"></app-child-a>`
})
export class ParentComponent {
  info = { message: 'Hello World' };
}

@Component({
  selector: 'app-child-a',
  template: `<app-child-b [data]="data"></app-child-b>`
})
export class ChildComponentA {
  @Input() data!: any;
}

@Component({
  selector: 'app-child-b',
  template: `<app-child-c [data]="data"></app-child-c>`
})
export class ChildComponentB {
  @Input() data!: any;
}

@Component({
  selector: 'app-child-c',
  template: `<p>{{ data.message }}</p>`
})
export class ChildComponentC {
  @Input() data!: any;
}
\end{lstlisting}

In Listing~\ref{lst:prop-drilling}, the \texttt{data} object defined in the \texttt{ParentComponent} must traverse multiple intermediate components before reaching its actual consumer, \texttt{ChildComponentC}. Furthermore, the inverse can occur when a deeply nested child component needs to emit an event to its ancestor. In this case, multiple intermediary components must define \texttt{@Output()} event bindings to propagate the emitted event upward in the hierarchy.

\subsubsection{Large file}\label{subsubsec:large-file}

In \textsc{Angular}, large files commonly emerge when developers embed templates directly within component classes using inline templates, combine multiple service implementations, or co-locate related but distinct functionalities without proper separation. This anti-pattern violates the principle of single responsibility at the file level, creating monolithic structures that become increasingly difficult to navigate and modify as the codebase evolves.

\begin{lstlisting}[language=TypeScript, caption=Large Angular file, label={lst:large-file-multiple}]
@Component({ selector: 'app-header', template: `<h1>{{ title }}</h1>` })
export class HeaderComponent { title = 'Header'; }

@Component({ selector: 'app-footer', template: `<footer>Footer</footer>` })
export class FooterComponent {}

@Injectable({ providedIn: 'root' })
export class AnalyticsService {
  log(event: string) { console.log(event); }
}

@Injectable({ providedIn: 'root' })
export class UserService {
  getUser() { return { name: 'Alice' }; }
}

// ...
\end{lstlisting}

Listing~\ref{lst:large-file-multiple} exemplifies a large file that aggregates multiple components and services within a single file.

\subsubsection{Duplicated component}\label{subsubsec:duplicated-component}

\textsc{Duplicated Component} manifests when multiple \textsc{Angular} components within a project share highly similar or identical structure, logic, or functionality. Consequently, such duplication leads to redundancy in the codebase. Furthermore, the presence of duplicated components suggests insufficient abstraction and poor reuse of common functionalities.

\begin{lstlisting}[language=TypeScript, caption=Duplicated Angular components, label={lst:duplicated-component}]
@Component({
    selector: 'app-user-card',
    template: `<div>{{ user.name }} - {{ user.email }}</div>`
})
export class UserCardComponent {
    @Input() user: { 
        name: string; 
        email: string 
    } = { name: '', email: '' };
}
@Component({
    selector: 'app-admin-card',
    template: `<div>{{ admin.name }} - {{ admin.email }}</div>`
})
export class AdminCardComponent {
    @Input() admin: { 
        name: string; 
        email: string
    } = { name: '', email: '' };
}
\end{lstlisting}

Listing~\ref{lst:duplicated-component} illustrates two components, \texttt{UserCardComponent} and \texttt{AdminCardComponent}, that exhibit highly similar structure and logic.

\subsection{(\textbf{RQ$_2$}) Where are \textsc{Angular} code smells discussed by the developer community?}\label{subsec:rq2}

We classified the sources in 7 categories, which are summarized in Table~\ref{tab:source-categories}. It shows the number of sources reviewed for each category. The most encountered types of sources are blog and forum posts. This predominance may indicate that most of the knowledge about front-end code smells is community-driven. Although informal sources may introduce subjectivity, they often capture practical issues that are overlooked by formal documentation or the literature.

\begin{table}[ht!] \small
    \caption{Categories of reviewed sources}
    \label{tab:source-categories}
    \begin{tabular}{lcc}
        \toprule
        \textbf{Category} & \textbf{\textit{Selected}} \\
        \midrule
        Personal Blog & 21 \\
        Forum Post & 15 \\
        Corporate Blog & 10 \\
        Course Material & 5 \\
        Video & 3 \\
        Q\&A & 2 \\
        Angular Documentation & 2 \\
        \bottomrule
    \end{tabular}
\end{table}

Among the analyzed materials, certain sources contributed more substantially to the identification of code smells. In particular, a Reddit post discussing anti-patterns in \textsc{Angular} projects comprised 53 comments and resulted in the identification of 10 distinct code smells. This finding underscores the relevance of community-driven discussions as valuable repositories of collective experiential knowledge regarding recurrent design and implementation issues.

\subsection{(\textbf{RQ$_3$}) Which \textsc{React}-specific code smells are also discussed by \textsc{Angular} practitioners?}\label{subsec:rq3}

Thematic analysis revealed that several \textsc{React}-specific smells are also discussed by \textsc{Angular} developers in grey literature. These smells were grouped under the \textit{cross-framework} category, as they represent design issues not exclusive to React, but rather to component-based front-end architectures more broadly. Table~\ref{tab:cross-framework-code-smells} summarizes these smells.

\begin{table} \small
    \caption{Cross-framework code smells}
    \label{tab:cross-framework-code-smells}
    \begin{tabular}{lcp{3cm}}
         \toprule
            \textbf{Code smells} & \textbf{\textit{Identified?}} \\
         \midrule
            Large Component & \checkmark \\
            Direct DOM Manipulation & \checkmark \\
            Inheritance Instead of Composition & \checkmark \\
            Too Many Inputs & \checkmark \\
            Prop Drilling & \checkmark \\
            Large File & \checkmark \\
            Duplicated Component & \checkmark \\
            Force Update & - \\
            Props in Initial State & - \\
            Uncontrolled Components & - \\
            JSX Outside Render Method & - \\
            Low Cohesion & - \\
         \bottomrule
    \end{tabular}
\end{table}

Several code smells originally identified in \textsc{React} were also prevalent in Angular grey literature, sometimes under different names. For example, \textsc{Too Many Props} in \textsc{React} corresponds closely to \textsc{Too Many Inputs} in \textsc{Angular}, both describing components overwhelmed with input bindings. Similarly, \textsc{Prop Drilling} is discussed as deep input/output chaining.

The presence of \textsc{Large File} and \textsc{Duplicated Components} further highlights common issues around code organization and reuse that transcend framework boundaries. Conversely, several \textsc{React}-specific smells related to its unique rendering model and syntax, such as \textsc{Force Update}, \textsc{Props in Initial State}, \textsc{Uncontrolled Components}, and \textsc{JSX Outside Render Method}, were absent in Angular literature.

\subsection{(\textbf{RQ$_4$}) How can existing code smells tools for \textsc{React} be extended to support the detection of \textsc{Angular} code smells?}\label{subsec:rq4}

The extension of existing detection tools required adapting the analysis process to accommodate the syntactic and structural differences between front-end frameworks. Although \textsc{ReactSniffer} and \textsc{SniffTSX} already operate in an integrated manner, their detectors were designed primarily for \textsc{React} and \textsc{TypeScript} projects. To enable multi-framework support, the execution pipeline was restructured to selectively activate detectors according to the framework identified in the target project. In this configuration, the \textsc{ReactSniffer} module remains exclusive to \textsc{React} analyses, while the \textsc{SniffTSX} detectors are executed in both \textsc{React}-based projects using \textsc{TypeScript} and in \textsc{Angular} projects.

Although several code smells are conceptually shared across frameworks (e.g., \textsc{Large Component}, \textsc{Too Many Props}), their syntactic realizations vary substantially. For example, an Angular component is declared as a class annotated with \texttt{@Component}, whereas a React component is defined either as a functional component or a class extending \texttt{React.Component}. Consequently, detectors cannot be directly reused across frameworks, as the underlying AST structures differ. Each cross-framework smell therefore requires framework-specific implementations that share a common detection logic but rely on distinct syntactic patterns.

%To support this, the proposed architecture adopts a modular design based on a unified detector interface (see Listing~\ref{lst:cross-framework-interface}). Concrete implementations specialize this interface for each framework, enabling reuse of conceptual logic while maintaining syntactic accuracy. For instance, a generic \textsc{Large Component} detector specifies the detection thresholds, whereas its React and Angular implementations process JSX elements or decorator-based components, respectively.

%\begin{lstlisting}[caption=Example of a cross-framework detector interface, label={lst:cross-framework-interface}]
%type AST = {
%    traverse: (visitor: Visitor) => void
%}

%interface DetectorContext {
%    framework: "react" | "angular"
%    ast: AST
%}

%type Smell {
%    start?: number
%    end?: number
%    path: string
%}

%type Detector = (context: DetectorContext) => Smell[]
%\end{lstlisting}

This modularization allows that each detector to be implemented as a pure function following a consistent signature. The \texttt{AST} type defines only the essential traversal capability, while its concrete implementation is provided by the parser associated with each framework. This design enables the developing and extension of the tool, allowing future integration of additional frameworks without altering the core architecture.

For instance, the detection of \textsc{Inheritance Instead of Composition} traverses the classes to identify occurrences of class inheritance through the \texttt{extends} keyword. When a component class inherits from another class, the detector flags the instance as a violation. In addition, the prototype version of the tool implements four other analyzers: \textsc{Overusing Any Type}, \textsc{Large Component}, \textsc{Large File}, and \textsc{Too Many Inputs}. These detectors encompass both \textsc{Framework-specific} and \textsc{Cross-framework} smells, ensuring that the architecture can be incrementally expanded to support the full catalog in future work.

\subsection{(\textbf{RQ$_5$}) How accurate and effective is the code smell detection tool in identifying code smells in open-source \textsc{Angular} projects?}\label{subsec:rq5}

The evaluation follows a multi-step procedure that involves selecting real-world projects, constructing and preparing a representative dataset, and applying the proposed tool to this dataset. The details of each step in the evaluation process are described in the following subsections.

\subsubsection{Project Selection}\label{subsec:project-selection}

To construct a representative set of real-world \textsc{Angular} applications, we selected ten publicly available open-source projects from GitHub. The selection process prioritized repositories with high community engagement, measured primarily through the number of stars, combined with explicit and consistent use of \textsc{Angular}. Repositories were first identified through the GitHub search and each candidate project was manually inspected to confirm the presence of Angular components.

The resulting set includes applications and libraries from diverse domains, including API clients, UI component frameworks, automation tools, and developer utilities, supporting an assessment of the proposed tool across heterogeneous and realistic usage scenarios. %Table~\ref{tab:angular-projects} summarizes the selected repositories.

%\vspace{0.3cm}
%\begin{table}[ht!] \small
%    \centering
%    \caption{Selected \textsc{Angular} projects used in the evaluation}
%    \label{tab:angular-projects}
%    \begin{tabular}{l}
%        \toprule
%        \textbf{Repository} \\
%        \midrule
%            angular/components \\
%            bitwarden/clients \\
%           igniteui/igniteui-angular \\
%            spartan-ng/spartan \\
%            taiga-family/taiga-ui \\
%            primefaces/primeng \\
%            ngx-charts \\
%            ng-bootstrap \\
%            Chocobozzz/PeerTube\\
%            NG-ZORRO/ng-zorro-antd \\
%        \bottomrule
%    \end{tabular}
%\end{table}

\subsubsection{Dataset Construction}\label{subsec:dataset-construction}

Because no publicly available dataset covers the code smells targeted by the proposed tool we constructed a labeled dataset to support the evaluation. Using the previously selected \textsc{Angular} projects, we manually inspected the source code to identify concrete and unambiguous instances of each smell. This manual process ensured that only representative examples aligned with the smell definitions were included.

For each supported smell type, 30 occurrences were selected, resulting in a dataset containing 150 instances in total. Then, each occurrence was manually refactored to eliminate the smell while preserving the component’s behavior. These refactored versions enable verification that the tool does not raise false positives when the underlying issue has been removed. Running the tool on both the original and refactored variants provides the basis for computing standard classification metrics, including Accuracy, Precision, Recall, and F1-score.

\subsubsection{Results}

Overall, the results indicate consistently high detection performance across all evaluated code smells. Accuracy values exceed 0.88 for every smell, while F1-scores range from 0.896 to 1.000, demonstrating the effectiveness of the proposed approach in identifying multiple types of code smells. Despite these strong results, variations across the evaluated metrics suggest that detection effectiveness is not uniform and is influenced by the specific characteristics of each smell. A detailed summary of these results is presented in Table~\ref{tab:detection-performance}.

\begin{table}[ht!] \scriptsize
\centering
\caption{Detection performance across different code smells}
\label{tab:detection-performance}
\begin{tabular}{lcccc}
\hline
\textbf{Code Smell} & \textbf{Accuracy} & \textbf{Precision} & \textbf{Recall} & \textbf{F1-score} \\
\hline
Overusing Any Type & 1.00 & 1.00 & 1.00 & 1.00 \\
Large Component   & 0.88 & 0.81 & 1.00 & 0.89 \\
Large File        & 0.95 & 0.90 & 1.00 & 0.95 \\
Inheritance Instead of Composition & 1.00 & 1.00 & 1.00 & 1.00 \\
Too Many Inputs & 1.00 & 1.00 & 1.00 & 1.00 \\
\hline
\end{tabular}
\end{table}

The Overusing Any Type smell achieves perfect performance across all evaluation metrics. This outcome suggests that the excessive use of the \texttt{any} type is characterized by clear and unambiguous syntactic patterns, which can be precisely identified through static analysis of the abstract syntax tree. As a result, the detector produces neither false positives nor false negatives for this smell in the evaluated dataset.

In contrast, the Large Component and Large File smells exhibit slightly lower detection performance, particularly in terms of precision, despite achieving a recall of 1.00. This indicates that all relevant instances were detected. However, this result was obtained with an increased number of false positives.

One recurring source of false positives observed in the proposed tool concerns size-related smells, such as Large Component and Large File. The detection strategy for these smells relies on absolute thresholds based on lines of code. Consequently, components or files exceeding a predefined size limit are flagged regardless of their internal organization, cohesion, or degree of modularization.

This behavior becomes evident in cases where components are conceptually well-structured but remain large due to domain complexity or legitimate functional aggregation. Listing~\ref{lst:modular-component} illustrates an example of a component that is flagged as a Large Component despite being modularized and cohesive.

\begin{lstlisting}[language=TypeScript, caption={Example of a modularized Angular component flagged as Large Component}, label={lst:modular-component}]
@Component({
  selector: 'app-dashboard',
  templateUrl: './dashboard.component.html'
})
export class DashboardComponent {

  loadUserData() { /* related logic */ }

  loadStatistics() { /* related logic */ }

  handleFilters() { /* related logic */ }

  updateViewState() { /* related logic */ }

  exportReport() { /* related logic */ }

}
\end{lstlisting}

Although the methods shown in Listing~\ref{lst:modular-component} are functionally related and collectively represent a cohesive responsibility, the component may still exceed the predefined line threshold. As a result, it is classified as a smell instance even after refactoring efforts aimed at improving modularity and separation of concerns.

\section{Discussion}

Code smells have been extensively investigated. However, their manifestations and implications can vary according to the architectural conventions and change detection mechanisms of each framework. In the context of \textsc{Angular}, our study identified a set of code smells that encompass both traditional object-oriented problems, such as \textsc{Inheritance Instead of Composition}, and framework-specific issues, such as \textsc{Inefficient Method Binding}.

A relevant observation emerging from our analysis concerns the conceptual correspondence between several of the identified Angular smells and those previously reported in other front-end frameworks. For instance, \textsc{Large Component} and \textsc{Too Many Inputs} correspond to similar modularity and responsibility issues described in \textsc{React}-based systems, such as \textsc{Large Component} and \textsc{Too Many Props}. These similarities suggest that a subset of design and maintainability issues is not restricted to a particular technology, but rather transcends framework boundaries.

%Another important aspect relates to the shared technological foundation of modern front-end frameworks. Since both \textsc{React} and \textsc{Angular} rely on TypeScript, several categories of anomalies, such as \textit{Overusing Any Type}, can be analyzed through a unified static analysis pipeline, independent of specific framework semantics. This observation indicates the feasibility of developing extensible detection tools capable of identifying both framework-specific and framework-agnostic smells.

Furthermore, the systematic identification of such smells can support the improvement of front-end development practices. By making design flaws explicit, smell detection tools can encourage developers to adopt refactoring and modularization strategies that improve code maintainability and readability. Integrating these detection mechanisms into IDEs would enable real-time feedback, allowing developers to recognize design anomalies as they code and to apply corrective measures proactively~\cite{Albuquerque2023IDDetection}. This continuous feedback loop can reduce technical debt accumulation and promote more sustainable development practices~\cite{Jacinto2021TechDebtCodeSmells}.

Beyond their immediate benefits for code quality, smell detection tools also hold pedagogical potential. When applied in educational or open-source contexts, these tools can foster experiential learning by allowing developers to explore, detect, and refactor smells in real-world scenarios~\cite{Eman2024RefactoringClassroom}. Such practice-oriented learning contributes to understanding software design principles, encourages community-driven quality improvements, and promotes the dissemination of maintainability-oriented development habits across the front-end ecosystem.

Finally, this study provides initial evidence that many of the identified smells in \textsc{Angular}-based systems can be generalized to other component-based frameworks, including \textsc{Vue} and \textsc{Svelte}. Several smells, notably \textit{Large Component}, \textit{Direct DOM Manipulation}, \textit{Duplicated Component}, etc., are likely to emerge in similar contexts. A broader investigation encompassing multiple frameworks would contribute to the consolidation of a unified catalog of front-end code smells.

\section{Related Work}

Research on object-oriented languages, particularly Java~\cite{Pereira2022SLRSmells, Martins2020SmellCo-ocurrences, Santana2024CodeStability, zakeri2023datasets} and C\#~\cite{Sharma2017SmellsCSharp, Partha2024xNose, sharma2021DeepDirectLearning}, has been extensive, focusing on traditional design issues. Some studies have expanded this scope to dynamically typed and scripting languages, such as Python~\cite{Boloori2025DpyPython} and JavaScript~\cite{Fard2013JsNose, Saboury2017EmpiricalJsSmells}, where the flexibility of the language often introduces distinct categories of anti-patterns.

Studies have also explored smell in emerging languages like Kotlin~\cite{Gupta2022Kotlin, Gois2019KotlinAndroid, Novendra2024KotlinDetection} and Elixir~\cite{Vegi2023CatalogElixir, Vegi2025RefactoringElixir}, underscoring how language-specific abstractions and functional constructs can reshape traditional notions of design quality. Beyond the language-level analyses, systematic reviews and empirical studies have examined smells in various software domains, including web systems~\cite{bessghaier2020webapps, Aniche2018MVC, Fawareh2024WebApps}, mobile applications~\cite{Fawad2025RefactoringAndroid, Prestat2024DynAMICSAndroid, Rahkema2020iOSSmells}, SQL~\cite{Biruk2020SQLCodeSmells}, and MongoDB~\cite{Boris2024Mongo}.

Research has investigated the perception and relevance of code smells from the developer's perspective. Surveys and replicated studies have explored whether developers consider certain smells meaningful and how frequently they act to resolve them~\cite{Hozano2018DevelopersCodeSmells, Tahir2020DevelopersStackExchange}. Complementary research has leveraged machine learning techniques to predict or detect code smells, including ensemble models, stacking classifiers, and deep learning approaches, often achieving high accuracy on large datasets~\cite{Esraa2025SmellML, ElJammal2025SmellPrediction}. Additionally, the emergence of large language models (LLMs) for software engineering has motivated studies on data smells in coding datasets~\cite{Alessandro2025RefactoringLLM, Antonio2025SmellsCodingTasks}.

Vegi et al.~\cite{Vegi2022CodeSmellsElixir} has investigated code smells in emerging and functional programming contexts. They conducted a grey literature review to identify code smells specific to Elixir. This study demonstrates that even languages with paradigms distinct from traditional object-oriented models exhibit recurring quality issues, emphasizing the importance of language and domain-specific investigations when analyzing bad software design practices.

In the context of JavaScript frameworks, Ferreira and Valente~\cite{Ferreira2023ReactSmells} identified twelve \textsc{React}-related code smells through a combination of grey literature review and interviews with professional developers, and a prototype tool, \textsc{ReactSniffer}, to automatically detect these smells. They also proposed examined the refactoring practices applied by developers to \textsc{React}~\cite{Ferreira2024ReactRefactoring}. By manually analyzing 320 refactoring commits, they cataloged 69 distinct operations, including \textsc{React}-specific refactorings, adaptations of traditional refactorings, and JavaScript/CSS-specific operations.

In a subsequent investigation, Nunes et al. \cite{maykonReactTypeScript} explored code smells in \textsc{React} applications written in TypeScript. The authors conducted a grey literature review, complemented by interviews and a community survey, to identify common quality issues specific to this context. They also developed \textsc{SniffTSX}, an extension of \textsc{ReactSniffer}. The tool was evaluated using a dataset of labeled instances, achieving high performance metrics.

\section{Threats to validity}\label{sec:threats-to-validity}

In this section, we discuss threats to the validity of our results \cite{wohlin2012experimentation}.

\textbf{Construct Validity}: Given that our source material consists of grey literature, which is not peer-reviewed, the terminology used to describe software quality issues is not always applied consistently. To mitigate this, we carefully reviewed each document to ensure that the identified issues align with commonly accepted definitions of code smells in the academic literature.

\textbf{Conclusion Validity}: Since our analysis is qualitative and based on manual classification, there is a risk of subjective interpretation. To reduce this threat, we adopted an iterative review process and maintained traceability between each smell and its respective source.

\textbf{Internal Validity}: One potential threat lies in the possibility of overlooking relevant grey literature sources. Although we followed systematic search procedures and explored multiple content platforms, we cannot guarantee that all relevant sources were captured. To minimize this, we documented our inclusion criteria and cross-validated results among the authors.

\textbf{External Validity}: As our study focuses exclusively on \textsc{Angular} and relies on grey literature, the findings may not be fully representative of industrial practice or directly applicable to other frameworks. Moreover, the relevance of specific smells may vary depending on the scale and context of the application. Despite these limitations, we believe that the identified smells reflect real-world concerns expressed by \textsc{Angular} practitioners and can serve as a foundation for future research and tool development.

\section{Conclusion and future work}

In this study, we investigated the presence of code smells in \textsc{Angular} applications by analyzing sources from grey literature. We identified 11 distinct smells reported by practitioners, six of which are also observed in \textsc{React}, indicating that several design problems transcend specific front-end frameworks. Based on these findings, we developed a prototype tool that automates the detection of six code smells in \textsc{Angular} projects and described the defining characteristics of each smell to support developers in their identification and mitigation.

The experimental evaluation demonstrates that the proposed detectors achieve consistently high detection performance. As summarized in Table~\ref{tab:detection-performance}, accuracy values remain above 0.88 for all implemented smells, while F1-scores range from 0.89 to 1.00. These results indicate that the proposed approach is effective in identifying multiple types of code smells, although variations across metrics suggest that detection effectiveness depends on the structural characteristics of each smell.

As future work, we intend to incorporate additional detectors into the tool, expanding its coverage to include a broader set of smells identified in the literature. Second, we aim to triangulate our findings with empirical analyses of real-world \textsc{Angular} codebases to better assess the prevalence and practical impact of the identified smells. Additionally, we plan to further evaluate the tool in diverse development contexts, including projects that combine multiple front-end frameworks such as \textsc{React} and \textsc{Angular}, and to enhance its reporting capabilities.

Finally, we envision extending the proposed methodology to other front-end frameworks, enabling the identification of both framework-specific and cross-framework code smells. This line of work contributes toward the construction of a comprehensive catalog of front-end code smells and the development of automated tools to support practitioners in improving the quality and maintainability of modern web applications.

\section*{Artifacts Availability}

All artifacts, including datasets, scripts, and documentation
from this study are available at \url{https://github.com/maykongsn/angular-code-smells}.

\section*{Acknowledgments}
This research was supported by CAPES - Finance Code 001; FAPESB grant PIE0002/2022; CNPq grants 403304/2025-3, 315840/2023-4 and 403361/2023-0.

\bibliographystyle{ACM-Reference-Format}
\bibliography{references}

\end{document}